\pdfoutput=1

\documentclass[12pt,a4paper]{article}

\usepackage{ifthen} 
\newboolean{pdflatex}
\setboolean{pdflatex}{true} 

\newboolean{articletitles}
\setboolean{articletitles}{true} 

\newboolean{uprightparticles}
\setboolean{uprightparticles}{false} 

\newboolean{inbibliography}
\setboolean{inbibliography}{false} 

\usepackage{microtype}
\usepackage{lineno}  
\usepackage{xspace} 

\usepackage{graphicx}  
\usepackage{color}
\usepackage{colortbl}
\graphicspath{{./figs/}} 

\usepackage{amsmath} 
\usepackage{amssymb}
\usepackage{amsfonts}
\usepackage{upgreek} 

\newcommand*\patchAmsMathEnvironmentForLineno[1]{%
\expandafter\let\csname old#1\expandafter\endcsname\csname #1\endcsname
\expandafter\let\csname oldend#1\expandafter\endcsname\csname
end#1\endcsname
 \renewenvironment{#1}%
   {\linenomath\csname old#1\endcsname}%
   {\csname oldend#1\endcsname\endlinenomath}%
}
\newcommand*\patchBothAmsMathEnvironmentsForLineno[1]{%
  \patchAmsMathEnvironmentForLineno{#1}%
  \patchAmsMathEnvironmentForLineno{#1*}%
}
\AtBeginDocument{%
\patchBothAmsMathEnvironmentsForLineno{equation}%
\patchBothAmsMathEnvironmentsForLineno{align}%
\patchBothAmsMathEnvironmentsForLineno{flalign}%
\patchBothAmsMathEnvironmentsForLineno{alignat}%
\patchBothAmsMathEnvironmentsForLineno{gather}%
\patchBothAmsMathEnvironmentsForLineno{multline}%
}

\usepackage{hyperref}    
\usepackage[all]{hypcap} 

\def\lhcb {\mbox{LHCb}\xspace}
\def\ux85 {\mbox{UX85}\xspace}

\ifthenelse{\boolean{uprightparticles}}%
{

 \def\Ppi         {\ensuremath{\uppi}\xspace}

 \def\Pphi        {\ensuremath{\upphi}\xspace}

 \def\PDelta      {\ensuremath{\Delta}\xspace}                 
 \def\PXi      {\ensuremath{\Xi}\xspace}                 
 \def\PLambda      {\ensuremath{\Lambda}\xspace}                 
 \def\PSigma      {\ensuremath{\Sigma}\xspace}                 
 \def\POmega      {\ensuremath{\Omega}\xspace}                 
 \def\PUpsilon      {\ensuremath{\Upsilon}\xspace}                 
 
 \def\PB      {\ensuremath{\mathrm{B}}\xspace}                 
                  
 \def\PD      {\ensuremath{\mathrm{D}}\xspace}

 \def\PK      {\ensuremath{\mathrm{K}}\xspace}

 \def\Pb      {\ensuremath{\mathrm{b}}\xspace}                 
 \def\Pc      {\ensuremath{\mathrm{c}}\xspace}

 \def\Pi      {\ensuremath{\mathrm{i}}\xspace}

 \def\Ps      {\ensuremath{\mathrm{s}}\xspace}

}
{

 \def\Ppi         {\ensuremath{\pi}\xspace}

 \def\Pphi        {\ensuremath{\phi}\xspace}

 \mathchardef\PDelta="7101
 \mathchardef\PXi="7104
 \mathchardef\PLambda="7103
 \mathchardef\PSigma="7106
 \mathchardef\POmega="710A
 \mathchardef\PUpsilon="7107
                  
 \def\PB      {\ensuremath{B}\xspace}                 
                  
 \def\PD      {\ensuremath{D}\xspace}

 \def\PK      {\ensuremath{K}\xspace}

 \def\Pb      {\ensuremath{b}\xspace}                 
 \def\Pc      {\ensuremath{c}\xspace}

 \def\Pi      {\ensuremath{i}\xspace}

 \def\Ps      {\ensuremath{s}\xspace}

}

\def\squark    {\ensuremath{\Ps}\xspace}

\def\cquark    {\ensuremath{\Pc}\xspace}

\def\bquark    {\ensuremath{\Pb}\xspace}
\def\bquarkbar {\ensuremath{\overline \bquark}\xspace}
\def\bbbar     {\ensuremath{\bquark\bquarkbar}\xspace}

\def\pion  {\ensuremath{\Ppi}\xspace}
\def\piz   {\ensuremath{\pion^0}\xspace}

\def\pip   {\ensuremath{\pion^+}\xspace}
\def\pim   {\ensuremath{\pion^-}\xspace}

\def\kaon  {\ensuremath{\PK}\xspace}
  \def\Kbar  {\kern 0.2em\overline{\kern -0.2em \PK}{}\xspace}

\def\Kz    {\ensuremath{\kaon^0}\xspace}
\def\Kzb   {\ensuremath{\Kbar^0}\xspace}
\def\KzKzb {\ensuremath{\Kz \kern -0.16em \Kzb}\xspace}
\def\Kp    {\ensuremath{\kaon^+}\xspace}
\def\Km    {\ensuremath{\kaon^-}\xspace}

\def\KpKm  {\ensuremath{\Kp \kern -0.16em \Km}\xspace}

\def\Kstarz  {\ensuremath{\kaon^{*0}}\xspace}
\def\Kstarzb {\ensuremath{\Kbar^{*0}}\xspace}

  \def\Dbar    {\kern 0.2em\overline{\kern -0.2em \PD}{}\xspace}
\def\D       {\ensuremath{\PD}\xspace}

\def\Dz      {\ensuremath{\D^0}\xspace}
\def\Dzb     {\ensuremath{\Dbar^0}\xspace}
\def\DzDzb   {\ensuremath{\Dz {\kern -0.16em \Dzb}}\xspace}
\def\Dp      {\ensuremath{\D^+}\xspace}
\def\Dm      {\ensuremath{\D^-}\xspace}

\def\DpDm    {\ensuremath{\Dp {\kern -0.16em \Dm}}\xspace}

\def\Dstarzb {\ensuremath{\Dbar^{*0}}\xspace}
\def\Dstarp  {\ensuremath{\D^{*+}}\xspace}

\def\Dsp     {\ensuremath{\D^+_\squark}\xspace}

\def\B       {\ensuremath{\PB}\xspace}
\def\Bbar    {\ensuremath{\kern 0.18em\overline{\kern -0.18em \PB}{}}\xspace}

\def\Bz      {\ensuremath{\B^0}\xspace}

\def\Bu      {\ensuremath{\B^+}\xspace}

\def\Bp      {\ensuremath{\Bu}\xspace}

\def\Bd      {\ensuremath{\B^0}\xspace}
\def\Bs      {\ensuremath{\B^0_\squark}\xspace}

  \def\Y#1S{\ensuremath{\PUpsilon{(#1S)}}\xspace}

\def\L {\ensuremath{\PLambda}\xspace}
\def\Lbar {\ensuremath{\kern 0.1em\overline{\kern -0.1em\PLambda}}\xspace}

\def\Lb      {\ensuremath{\L^0_\bquark}\xspace}

\newcommand{\decay}[2]{\ensuremath{#1\!\to #2}\xspace}         

\def\to                 {\ensuremath{\rightarrow}\xspace}

\def\CP                {\ensuremath{C\!P}\xspace}

\newcommand{\betas}{\ensuremath{\beta_{\squark}}\xspace}

\def\BsToDphi     {\decay{\Bs}{\Dzb\phi}}
\def\BsToDKstar   {\decay{\Bs}{\Dzb\Kstarzb}}
\def\BdToDKstar   {\decay{\Bd}{\Dzb\Kstarz}}

\def\AT#1     {\ensuremath{A_{\mathrm{T}}^{#1}}\xspace}           

\def\C#1      {\ensuremath{\mathcal{C}_{#1}}\xspace}                       
\def\Cp#1     {\ensuremath{\mathcal{C}_{#1}^{'}}\xspace}                    
\def\Ceff#1   {\ensuremath{\mathcal{C}_{#1}^{\mathrm{(eff)}}}\xspace}        
\def\Cpeff#1  {\ensuremath{\mathcal{C}_{#1}^{'\mathrm{(eff)}}}\xspace}       
\def\Ope#1    {\ensuremath{\mathcal{O}_{#1}}\xspace}                       
\def\Opep#1   {\ensuremath{\mathcal{O}_{#1}^{'}}\xspace}                    

\newcommand{\tev}{\ifthenelse{\boolean{inbibliography}}{\ensuremath{~T\kern -0.05em eV}\xspace}{\ensuremath{\mathrm{\,Te\kern -0.1em V}}\xspace}}
\newcommand{\gev}{\ensuremath{\mathrm{\,Ge\kern -0.1em V}}\xspace}
\newcommand{\mev}{\ensuremath{\mathrm{\,Me\kern -0.1em V}}\xspace}
\newcommand{\kev}{\ensuremath{\mathrm{\,ke\kern -0.1em V}}\xspace}
\newcommand{\ev}{\ensuremath{\mathrm{\,e\kern -0.1em V}}\xspace}
\newcommand{\gevc}{\ensuremath{{\mathrm{\,Ge\kern -0.1em V\!/}c}}\xspace}
\newcommand{\mevc}{\ensuremath{{\mathrm{\,Me\kern -0.1em V\!/}c}}\xspace}
\newcommand{\gevcc}{\ensuremath{{\mathrm{\,Ge\kern -0.1em V\!/}c^2}}\xspace}
\newcommand{\gevgevcccc}{\ensuremath{{\mathrm{\,Ge\kern -0.1em V^2\!/}c^4}}\xspace}
\newcommand{\mevcc}{\ensuremath{{\mathrm{\,Me\kern -0.1em V\!/}c^2}}\xspace}

\def\mum  {\ensuremath{\,\upmu\rm m}\xspace}

\def\invfb   {\ensuremath{\mbox{\,fb}^{-1}}\xspace}

\newcommand{\chisq}{\ensuremath{\chi^2}\xspace}

\def\gsim{{~\raise.15em\hbox{$>$}\kern-.85em
          \lower.35em\hbox{$\sim$}~}\xspace}
\def\lsim{{~\raise.15em\hbox{$<$}\kern-.85em
          \lower.35em\hbox{$\sim$}~}\xspace}

\def\sPlot{\mbox{\em sPlot}}

\def\pt         {\mbox{$p_{\rm T}$}\xspace}

\def\evtgen     {\mbox{\textsc{EvtGen}}\xspace}
\def\pythia     {\mbox{\textsc{Pythia}}\xspace}

\def\geant      {\mbox{\textsc{Geant4}}\xspace}

\def\photos     {\mbox{\textsc{Photos}}\xspace}

\def\tell1  {TELL1\xspace}
\def\ukl1   {UKL1\xspace}

\newcommand{\eg}{\mbox{\itshape e.g.}\xspace}

\xspace

\usepackage{cite} 
\usepackage{mciteplus}

\usepackage{longtable} 

\begin{document}

\renewcommand{\thefootnote}{\fnsymbol{footnote}}
\setcounter{footnote}{1}


\begin{titlepage}
\pagenumbering{roman}

\vspace*{-1.5cm}
\centerline{\large EUROPEAN ORGANIZATION FOR NUCLEAR RESEARCH (CERN)}
\vspace*{1.5cm}
\hspace*{-0.5cm}
\begin{tabular*}{\linewidth}{lc@{\extracolsep{\fill}}r}
\ifthenelse{\boolean{pdflatex}}
{\vspace*{-2.7cm}\mbox{\!\!\!\includegraphics[width=.14\textwidth]{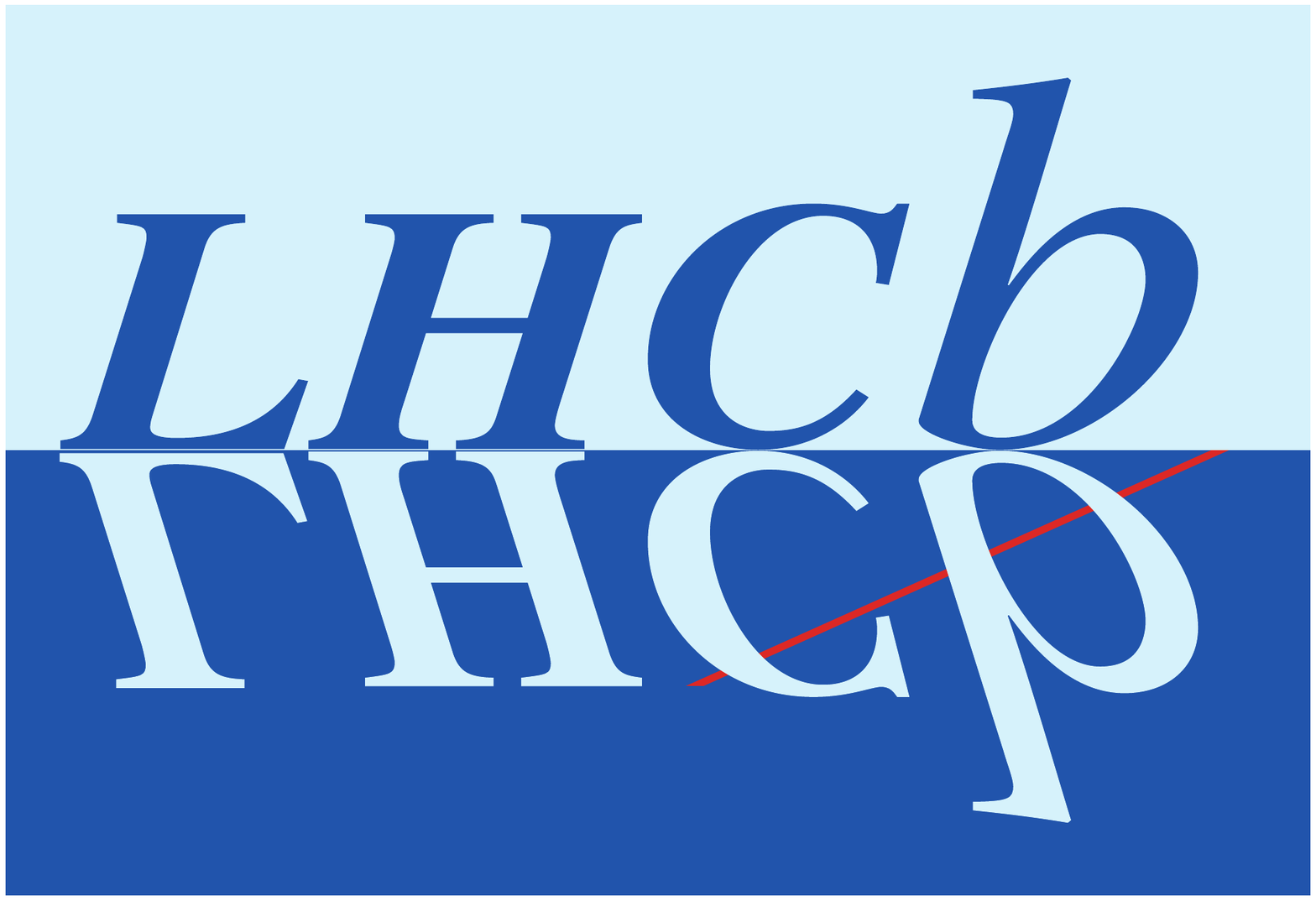}} & &}%
{\vspace*{-1.2cm}\mbox{\!\!\!\includegraphics[width=.12\textwidth]{lhcb-logo}} & &}%
\\
 & & CERN-PH-EP-2013-150 \\  
 & & LHCb-PAPER-2013-035 \\  
 & & 21 August 2013 \\ 
 & & \\
\end{tabular*}

\vspace*{3.0cm}

{\bf\boldmath\huge
\begin{center}
Observation of the decay $\Bs\to\Dzb\phi$   
\end{center}
}

\vspace*{1.5cm}

\begin{center}
The LHCb collaboration\footnote{Authors are listed on the following pages.}
\end{center}

\vspace{\fill}

\begin{abstract}
  \noindent
First observation of the decay \BsToDphi is reported using $pp$ collision data, 
corresponding to an integrated luminosity of 1.0\,\invfb, collected by the
LHCb experiment at a centre-of-mass energy of $7$\,TeV.
The significance of the signal is 6.5 standard deviations.
The branching fraction is measured relative to that of the decay \BsToDKstar to be
$$\frac{{\cal{B}}(\BsToDphi)}{{\cal{B}}(\BsToDKstar)} = 0.069 \pm 0.013 ~(\mathrm{stat}) \pm 0.007 ~(\mathrm{syst}).$$ 
The first measurement of 
the ratio of branching fractions for the decays \BsToDKstar and
\BdToDKstar is found to be
$$\frac{{\cal{B}}(\BsToDKstar)}{{\cal{B}}(\BdToDKstar)} = 7.8 \pm 0.7 ~(\mathrm{stat}) \pm 0.3 ~(\mathrm{syst}) \pm 0.6 ~(f_s/f_d),$$
where the last uncertainty is due to the ratio of the \Bs and \Bd fragmentation fractions.  

\end{abstract}
\vspace*{1.5cm}

\begin{center}
  Submitted to Phys. Lett. B
\end{center}

\vspace{\fill}

{\footnotesize 
\centerline{\copyright~CERN on behalf of the \lhcb collaboration, license \href{http://creativecommons.org/licenses/by/3.0/}{CC-BY-3.0}.}}
\vspace*{2mm}

\end{titlepage}


\newpage
\setcounter{page}{2}
\mbox{~}
\newpage

\centerline{\large\bf LHCb collaboration}
\begin{flushleft}
\small
R.~Aaij$^{40}$, 
B.~Adeva$^{36}$, 
M.~Adinolfi$^{45}$, 
C.~Adrover$^{6}$, 
A.~Affolder$^{51}$, 
Z.~Ajaltouni$^{5}$, 
J.~Albrecht$^{9}$, 
F.~Alessio$^{37}$, 
M.~Alexander$^{50}$, 
S.~Ali$^{40}$
% ...

{\footnotesize \it
% ...
$ ^{5}$Clermont Universit\'{e}, Universit\'{e} Blaise Pascal, CNRS/IN2P3, LPC, Clermont-Ferrand, France\\
$ ^{6}$CPPM, Aix-Marseille Universit\'{e}, CNRS/IN2P3, Marseille, France\\
% ...
$ ^{9}$Fakult\"{a}t Physik, Technische Universit\"{a}t Dortmund, Dortmund, Germany\\
% ...
$ ^{36}$Universidad de Santiago de Compostela, Santiago de Compostela, Spain\\
$ ^{37}$European Organization for Nuclear Research (CERN), Geneva, Switzerland\\
% ...
$ ^{40}$Nikhef National Institute for Subatomic Physics, Amsterdam, The Netherlands\\
% ...
$ ^{45}$H.H. Wills Physics Laboratory, University of Bristol, Bristol, United Kingdom\\
% ...
$ ^{50}$School of Physics and Astronomy, University of Glasgow, Glasgow, United Kingdom\\
$ ^{51}$Oliver Lodge Laboratory, University of Liverpool, Liverpool, United Kingdom\\
% ...
}
\end{flushleft}


\cleardoublepage


\renewcommand{\thefootnote}{\arabic{footnote}}
\setcounter{footnote}{0}



\pagestyle{plain} 
\setcounter{page}{1}
\pagenumbering{arabic}


\section{Introduction}
\label{sec:Introduction}

Measurements of the decay\footnote{The inclusion of charge conjugate processes is implied, unless otherwise stated.} \BsToDphi are of particular interest because they provide information that can be used to determine the CKM angles $\gamma\,\equiv\mathrm{arg}[-V_{ud}V_{ub}^*/(V_{cd}V_{cb}^*)]$ and $\beta_s\,\equiv\mathrm{arg}[-V_{ts}V_{tb}^*/(V_{cs}V_{cb}^*)]$ without theoretical uncertainties~\cite{bib:NandiLondon}. Knowledge of these \CP-violating phases is crucial to search for new sources of \CP violation and unravel subtle effects of physics beyond the Standard Model, which may appear in flavour-changing interactions. Their precise measurements are among the most important goals of flavour physics experiments.

To date, the angle $\gamma$ is the least well-determined angle of the Unitarity Triangle with an uncertainty of about $10^\circ$~\cite{Lees:2013zd, Trabelsi:2013uj, Aaij:2013zfa}. The current precision is dominated by measurements of time-integrated $\Bp\to DK^+$ decay rates, where $D$ indicates a superposition of \Dz and \Dzb decays to a common final state. In these decays, sensitivity to $\gamma$ arises from direct \CP violation in the interference between the $b\to c\bar{u}s$ and  $b\to u\bar{c}s$ tree-level amplitudes. As there are no loop contributions to the decay amplitudes, no theoretical uncertainties arise. The main limitation is due to the size of the data samples collected by the experiments. To improve on the precision, it is important to perform additional measurements from other channels with small theoretical uncertainties.

The large production cross-section of \Bs mesons in $pp$ collisions at the LHC opens new possibilities for measuring 
both $\gamma$ and $\beta_s$. For example, the decay
$\Bs\to D_s^{\pm}K^{\mp}$ is sensitive to $\gamma + 2\beta_s$ through measurements of time-dependent decay rates
~\cite{Aleksan:1991nh,Fleischer:2003yb};
although the determination of $\gamma$ from this mode requires an independent measurement of the mixing phase $\beta_s$. 

The decay \BsToDphi, first proposed in 1991 by Gronau and London for measuring $\gamma$~\cite{bib:GronauLondon}, 
can also probe $\beta_s$ via measurements of time-dependent decay rates.
Nandi and London have shown~\cite{bib:NandiLondon} that both $\gamma$ and $\betas$ can be determined without theoretical uncertainties and ambiguities, using the known sign of $\Delta \Gamma_s$, the decay-width difference between the two \Bs mass eigenstates~\cite{Aaij:2012eq}.

An alternative method to measure $\gamma$ using $\Bs\to D\phi$ decays was proposed in Refs.~\cite{bib:GGSSZ, bib:GGSZ},
where it was shown that $\gamma$ can be determined from time-integrated decay rates, 
in a similar way as from $\Bp\to D\Kp$ decays, even if $\Bs\to D\phi$ is not a self-tagged decay mode.
The only requirement for the determination is  
that a sufficient number of different $D$ final states are included in the measurement.
The time-integrated method does not require flavour-tagging, and 
hence makes optimal use of the statistical power of the large \bbbar production at LHC. 
An estimation of the sensitivity with this method
shows that the mode $\Bs\to D\phi$ has the potential to make a significant impact on the determination of $\gamma$
at \lhcb~\cite{bib:SRPubNote}. 

\begin{figure}[htp!]
  \begin{center}
    \includegraphics[width=0.35\linewidth]{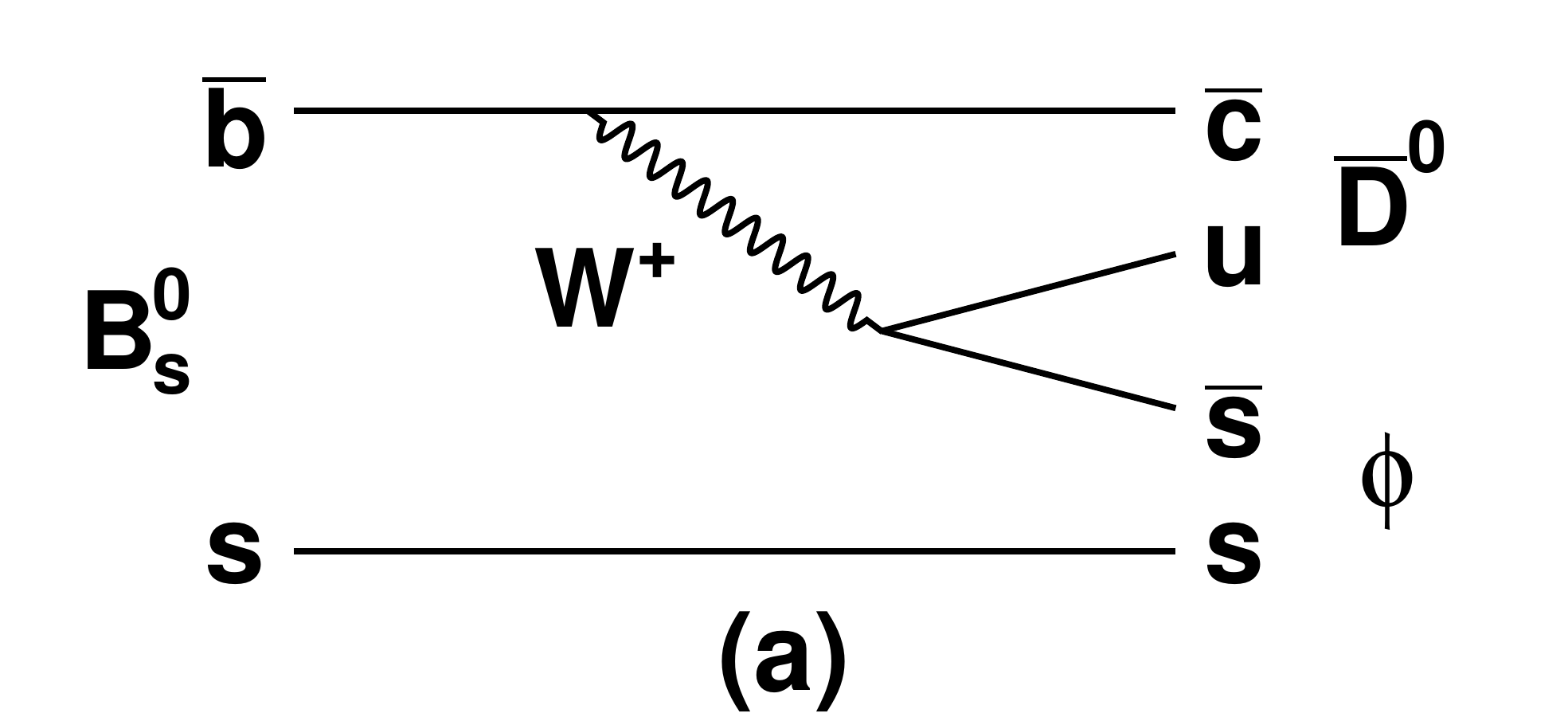}
    \includegraphics[width=0.35\linewidth]{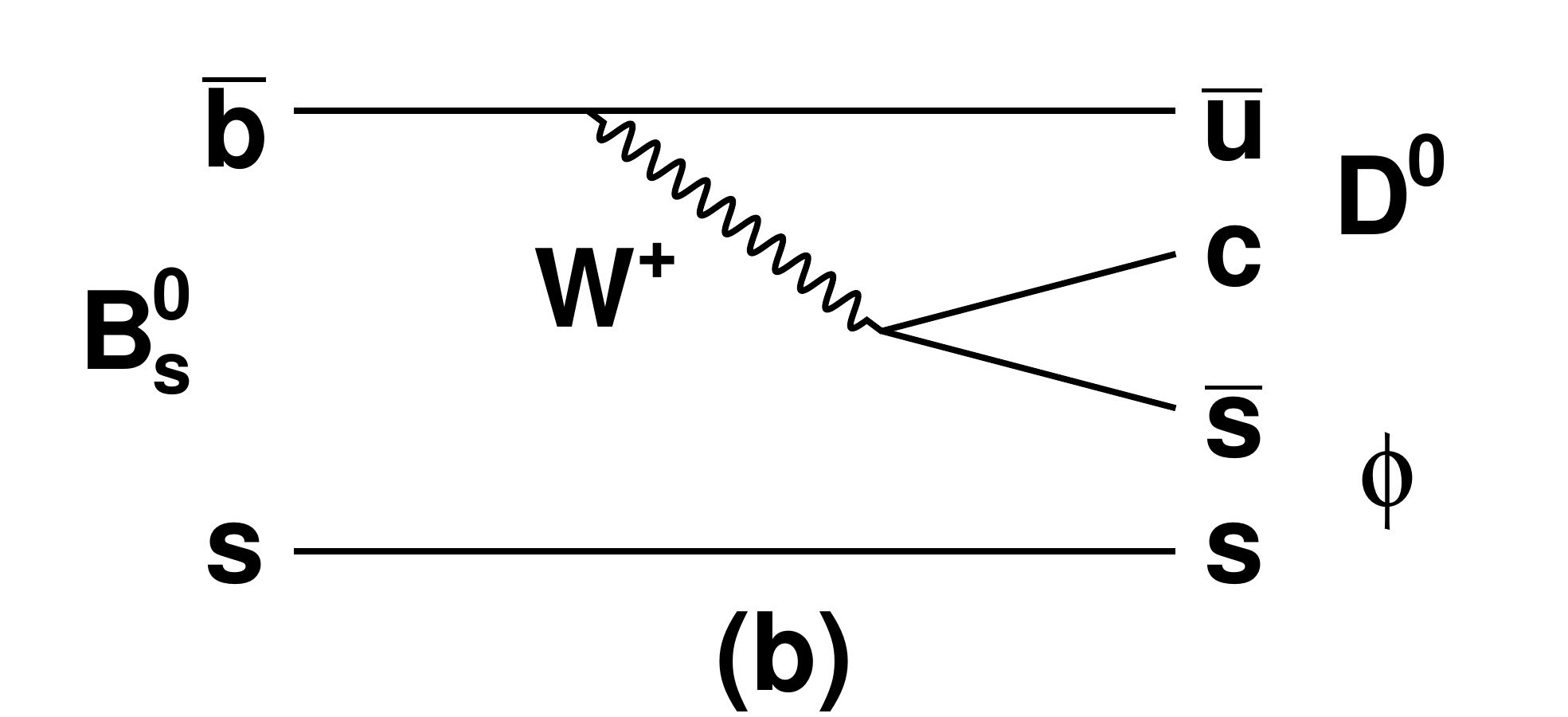}
    \includegraphics[width=0.35\linewidth]{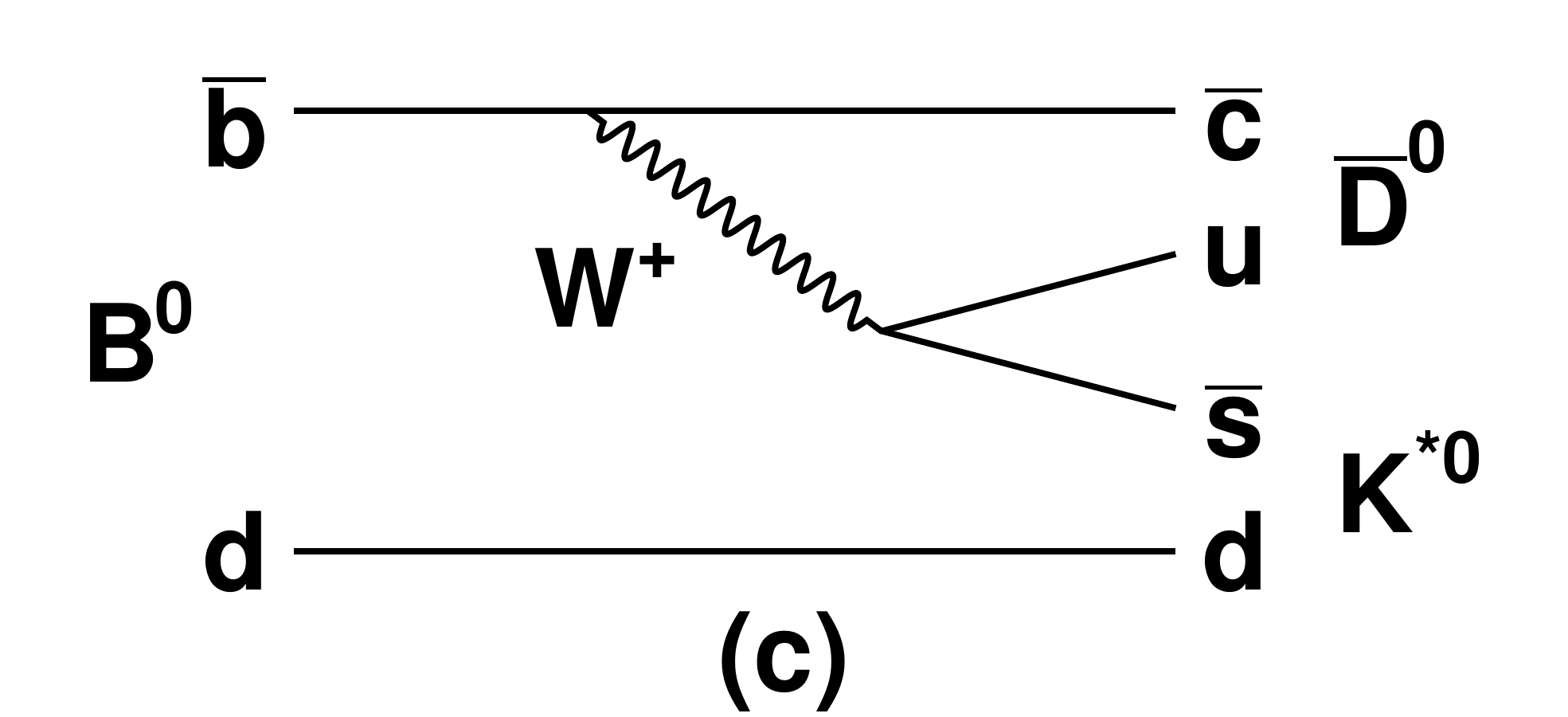}
    \includegraphics[width=0.35\linewidth]{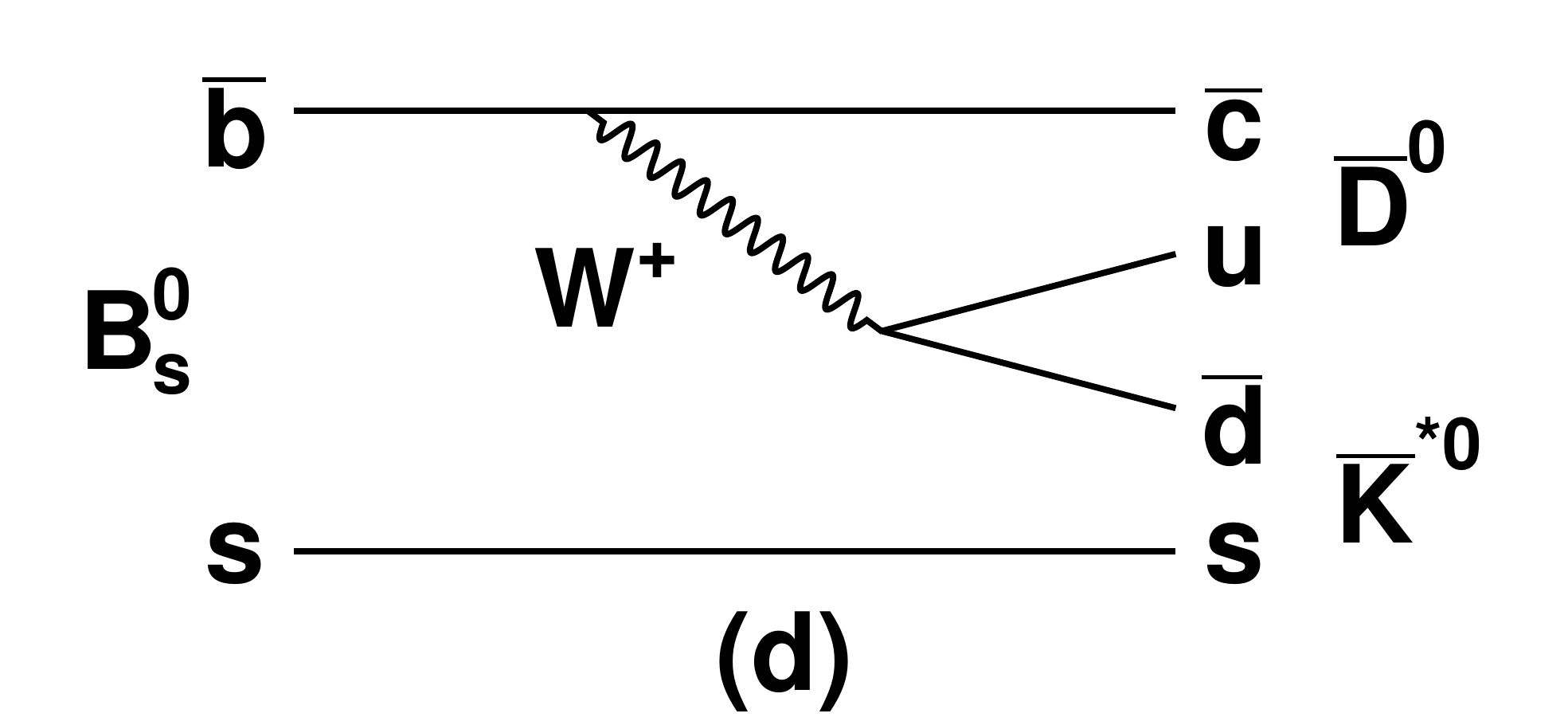}
  \end{center}
  \caption{
    \small 
Feynman diagrams for the following decays: (a) \BsToDphi; (b) $\Bs\to\Dz\Pphi$; (c) \BdToDKstar; and (d) \BsToDKstar.
The \BsToDphi and $\Bs\to\Dz\Pphi$ decay amplitudes interfere when $\Dzb$ and $\Dz$ decay to the same final state. 
    }
  \label{fig:Feyn}
\end{figure}
The observation of the \BsToDphi decay and the measurement of its branching fraction, described in this Letter, 
are the first steps towards a programme of \CP violation studies with this channel.
The branching fraction is measured relative to the topologically similar decay \BsToDKstar, that was previously 
observed by \lhcb~\cite{bib:LHCb-PAPER-2011-008}. In addition, the first measurement of the branching fraction of the \BsToDKstar decay relative to the \BdToDKstar decay is reported and used to improve on the knowledge of the branching fraction of the \BsToDKstar decay. The Feynman diagrams corresponding to the \BsToDphi and $\Bs\to \Dz\phi$ decay amplitudes 
are shown in Fig.~\ref{fig:Feyn}. 
The Feynman diagrams for the leading $b\to c$ amplitudes 
in \BsToDKstar and \BdToDKstar decays are also shown in Fig.~\ref{fig:Feyn}.
Since only $\Dz\to\Km\pip$ decays are considered in this study, 
all of the measured quantities for the \BsToDphi, \BsToDKstar, and \BdToDKstar channels
include contributions from the $\Bs\to\Dz\phi$, $\Bs\to\Dz\Kstarzb$, and $\Bd\to\Dz\Kstarz$ modes, respectively, through the doubly-Cabibbo-suppressed decay $\Dz\to\Kp\pim$. 

\section{Event selection}
\label{sec:Selection}

The study reported here is based on $pp$ collision data, corresponding to an integrated luminosity of  1.0\,\invfb, 
collected by the LHCb experiment at a centre-of-mass energy of 7\,TeV.
The \lhcb detector~\cite{Alves:2008zz} is a single-arm forward
spectrometer covering the \mbox{pseudorapidity} range $2<\eta <5$,
designed for the study of particles containing \bquark or \cquark
quarks. The detector includes a high-precision tracking system
consisting of a silicon-strip vertex detector surrounding the $pp$
interaction region, a large-area silicon-strip detector located
upstream of a dipole magnet with a bending power of about
$4{\rm\,Tm}$, and three stations of silicon-strip detectors and straw
drift tubes placed downstream.
The combined tracking system provides a momentum ($p$) measurement with
relative uncertainty that varies from 0.4\% at 5\gevc to 0.6\% at 100\gevc,
and impact parameter (IP) resolution of 20\mum for
tracks with large transverse momentum (\pt). Charged hadrons are identified
using two ring-imaging Cherenkov detectors~\cite{Adinolfi:2012an}. Photon, electron and
hadron candidates are identified by a calorimeter system consisting of
scintillating-pad and preshower detectors, an electromagnetic
calorimeter and a hadronic calorimeter. Muons are identified by a
system composed of alternating layers of iron and multiwire
proportional chambers~\cite{LHCb-DP-2012-002}.
The trigger~\cite{LHCb-DP-2012-004} consists of a
hardware stage, based on information from the calorimeter and muon
systems, followed by a software stage, which applies a full event
reconstruction.

Simulated signal samples and data control channels are used to optimise the selection criteria.
In the simulation, $pp$ collisions are generated using \pythia~6.4~\cite{Sjostrand:2006za} with a specific \lhcb
configuration~\cite{LHCb-PROC-2010-056}.  Decays of hadrons
are described by \evtgen~\cite{Lange:2001uf}, in which final state
radiation is generated using \photos~\cite{Golonka:2005pn}. The
interaction of the generated particles with the detector and its
response are implemented using the \geant
toolkit~\cite{Allison:2006ve, *Agostinelli:2002hh} as described in
Ref.~\cite{LHCb-PROC-2011-006}.

Selected events fulfill one of two hardware trigger requirements: either
a particle from the signal decay deposits enough energy
in the calorimeter system, or one of the particles in the event, not originating 
from the signal decay, fulfils any of the trigger requirements (\eg, 
events triggered by one or more particles coming from the decay of the other $B$ meson in the $pp\to b\bar{b} X$ event).  
The software trigger requires a two-, three- or four-track
secondary vertex with a large scalar sum of
the tracks \pt and significant displacement from the associated primary $pp$
interaction vertex~(PV). At least one track should have $\pt >
1.7\gevc$ and a value of $\chisq_{\rm IP} > 16$, where $\chisq_{\rm IP}$ is defined as the
difference between the \chisq of the PV reconstructed with and
without the considered particle. A multivariate algorithm identifies
secondary vertices consistent with the decay of a \bquark hadron.

Reconstructed tracks are selected with criteria on their $p$, \pt, track \chisq per degree of freedom, $\chisq_{\rm IP}$ and particle identification (PID). 
Tracks identified as muons are discarded. 

The \Dz mesons are reconstructed in the decay mode $\Dz\to \Km\pip$. Particle identification 
criteria used to select the daughters require
the difference between the log-likelihoods of the kaon and pion hypotheses ($\Delta LL_{K\pi}$) to be larger 
than 0 for the kaon and smaller than 4 for the pion. 
The \Dz meson $\chisq_{\rm IP}$ is required to be larger than 2 to separate 
mesons originating from a $B$ decay and those produced at the PV. In addition, 
for the $\Dzb\Kstarz$ ($\Dzb\Kstarzb$) final states, the charm meson flight distance 
with respect to the $B^0_{(s)}$ vertex is required to be larger than 0 with a significance of at least 2 standard deviations 
in order to suppress background from $B^0_{(s)}$ decays without an intermediate charm meson, such as 
the mode $\Bz\to\Km\pip\Kstarz$. There is no corresponding requirement in the $\Dzb\phi$ final state, since the charmless background 
is negligible. The \Dz candidates with invariant mass within $\pm 20\,\mevcc$ 
of the known mass~\cite{PDG2012} are retained.

The \Pphi mesons are reconstructed in the mode $\phi\to\Kp\Km$. The \pt of the kaon daughters is required to be 
larger than 350\,\mevc and the $\Delta LL_{K\pi}$ of both daughters to be larger than 3. Candidates are retained if their invariant mass is within $\pm 10\,\mevcc$ of the known \Pphi mass~\cite{PDG2012}.

The \Kstarz mesons are reconstructed in the mode $\Kstarz\to\Kp\pim$. The \pt of the kaon\,(pion) is required to be larger than 350\,(250)\,\mevc. In addition, to reduce the cross-feed from $\Bz\to \Dzb\rho^0$ 
and $\Bz\to \Dzb\Kp\Km$ decays, the $\Delta LL_{K\pi}$ of the kaon must be larger than 3 and that of the pion smaller than 3. Possible background from protons in the kaon sample, for example from the decay $\Lb\to\Dz p\pim$, is suppressed by selecting kaon candidates with a difference between the log-likelihoods of proton and kaon hypotheses, $\Delta LL_{pK}$, smaller than 10. 
Candidate \Kstarz mesons with invariant mass within $\pm 50\,\mevcc$ of the known mass~\cite{PDG2012} are kept.

Neutral $B$ meson candidates are formed from \Dzb and \Pphi(or \Kstarz) candidates, which are fitted to a common vertex with the \Dzb constrained to its known mass. In order to reduce contributions from non-resonant decays, $B^0_{(s)}\to \Dzb\Kp\Km$, $B^0_s\to \Dzb\Km\pip$, and $B^0\to \Dzb\Kp\pim$~\cite{bib:BDKK, Aaij:2013pua}, the absolute value of the cosine of the vector-daughter helicity angle ($\mathrm{cos}\,\theta_h$) is required to be larger than 0.4. This angle is defined between the momentum direction of the \Kp daughter in the $\phi\,(\Kstarz)$ frame, and the vector meson direction in the $B$ rest frame. Backgrounds from $\B^0_{(s)} \to D^\mp_{(s)}h^\pm$\,($h = \pi, K$) decays, are rejected by vetoing candidates 
with $\Kp\Km\pip$ ($\Km\pip\pip$  and $\Kp\Km\pip$) invariant mass within $\pm\,15\,\mevcc$ of the \Dsp (\Dp) meson known mass~\cite{PDG2012}. 

A boosted decision tree (BDT)~\cite{Breiman} suppresses the residual background.
Nine variables are input to the BDT: the decay vertex \chisq of the reconstructed $B^0_{(s)}$ and \Dz mesons; the $\chisq_{\rm IP}$ of the $B^0_{(s)}$, \Dz, \Pphi ($\Kstarz$) mesons, and of both the \Dz daughters; and the \pt of the \Dz and $\phi$ (\Kstarz) mesons. The BDT is optimised and tested using 
simulated signal events and events outside of the \Dz mass signal region 
 for background. Events with BDT response larger than 0.2 are retained, resulting in a rejection of 74\%
of the background, while retaining 84\% of the signal.
The working point maximises $N_\mathrm{s}/\sqrt{N_\mathrm{s} + N_\mathrm{b}}$. Here, $N_\mathrm{s}$ 
is the expected \BsToDphi signal yield, computed using simulated events and 
assuming that the branching fraction is equal to that of the \BdToDKstar decay (as expected under SU(3) flavour symmetry), 
and $N_\mathrm{b}$ is the background yield estimated using data events in the sidebands outside the \BsToDphi signal region ($\pm\,50$\mevcc around the \Bs known mass~\cite{PDG2012}). 
No multiple candidates are found for the $\Dzb\phi$ final state. 
The fraction of events with more than one candidate is 0.6\% in the $\Dzb\Kstarzb$ or $\Dzb\Kstarz$ 
invariant mass range of 5150--5600\,\mevcc, and the candidate retained is chosen randomly.

\section{Signal yield}
\label{sec:Fit}

Signal yields are determined with an unbinned maximum likelihood fit to the $\Dzb\phi$ and the sum of the $\Dzb\Kstarz$ and $\Dzb\Kstarzb$ invariant mass ($M$) distributions in the range $5150 < M < 5600$~\mevcc. The two samples are fitted simultaneously with a sum of probability density functions (PDFs) modelling signal and background contributions. 

The \Bs and \Bz signals are described by a modified Gaussian distribution of the form
\begin{equation}
f(M;\mu,\sigma,\alpha_L,\alpha_R) \propto \mathrm{exp}\left(\frac{-(M-\mu)^2}{2\sigma^2 + \alpha_{L,R}(M-\mu)^2}\right),
\end{equation}
where $\mu$ is the peak position, $\sigma$ the width, and $\alpha_{L}\,(M<\mu)$ and  $\alpha_{R}\,(M>\mu)$ parameterise the tails.
The width and the tail parameters depend on the final state, but are common to the \Bs and \Bz decays. 
The \Bz peak position and width are left free to vary in the fit with the difference between 
\Bs and \Bz peak positions fixed to the current world-average value~\cite{PDG2012}. The tail parameters
are fixed to values determined from simulated events and are considered among the sources of systematic uncertainty.
The recently observed decay $\Bz\to\Dzb\Kp\Km$~\cite{bib:BDKK} is expected to contribute to the $\Dzb\phi$ distribution and is modelled with the same modified Gaussian distribution, but with different peak position, as that used to describe the \BsToDphi decay.

Background from the $\Bz \to \Dzb\rho^0$ decay in the $\Dzb\Kstarz$ (or $\Dzb\Kstarzb$) final state can arise from misidentification of one of the pions from the $\rho^0\to\pip\pim$ decay as a kaon. The shape of this cross-feed contribution is modelled with a Crystal Ball function~\cite{Skwarnicki:1986xj} determined from simulated events. This background component is absent in the $\Bs\to\Dzb\phi$ mode, since the probability that both pions are misidentified as kaons and that their invariant mass is inside the narrow \Pphi mass window is negligible. For similar reasons, the cross-feed between $\Bz \to\Dzb\Kstarz$ and \BsToDphi decays is negligible.

The decay $\Bs\to\Dstarzb\Kstarzb$, where a \piz or photon from the \Dstarzb decay is not reconstructed, constitutes the main background contribution to the $\Dzb\Kstarzb$ final state below the \Bz mass. Similarly, the decay $\Bs\to\Dstarzb\phi$ is expected to contribute to the low-mass background in the $\Dzb\phi$ final state. These decays of a pseudoscalar to two vector mesons are modelled by a non-parametric 
PDF~\cite{bib:Kernel} determined from simulation. The mass shape depends on the unknown fraction of longitudinal polarisation, 
which is assumed to be identical for the two modes and is treated as an additional free parameter in the fit.

The remaining combinatorial background is described by a linear function, with a common slope for the two considered final states, left free to vary in the fit. 

\begin{figure}[tb!]
  \begin{center}
    \includegraphics[width=0.49\linewidth]{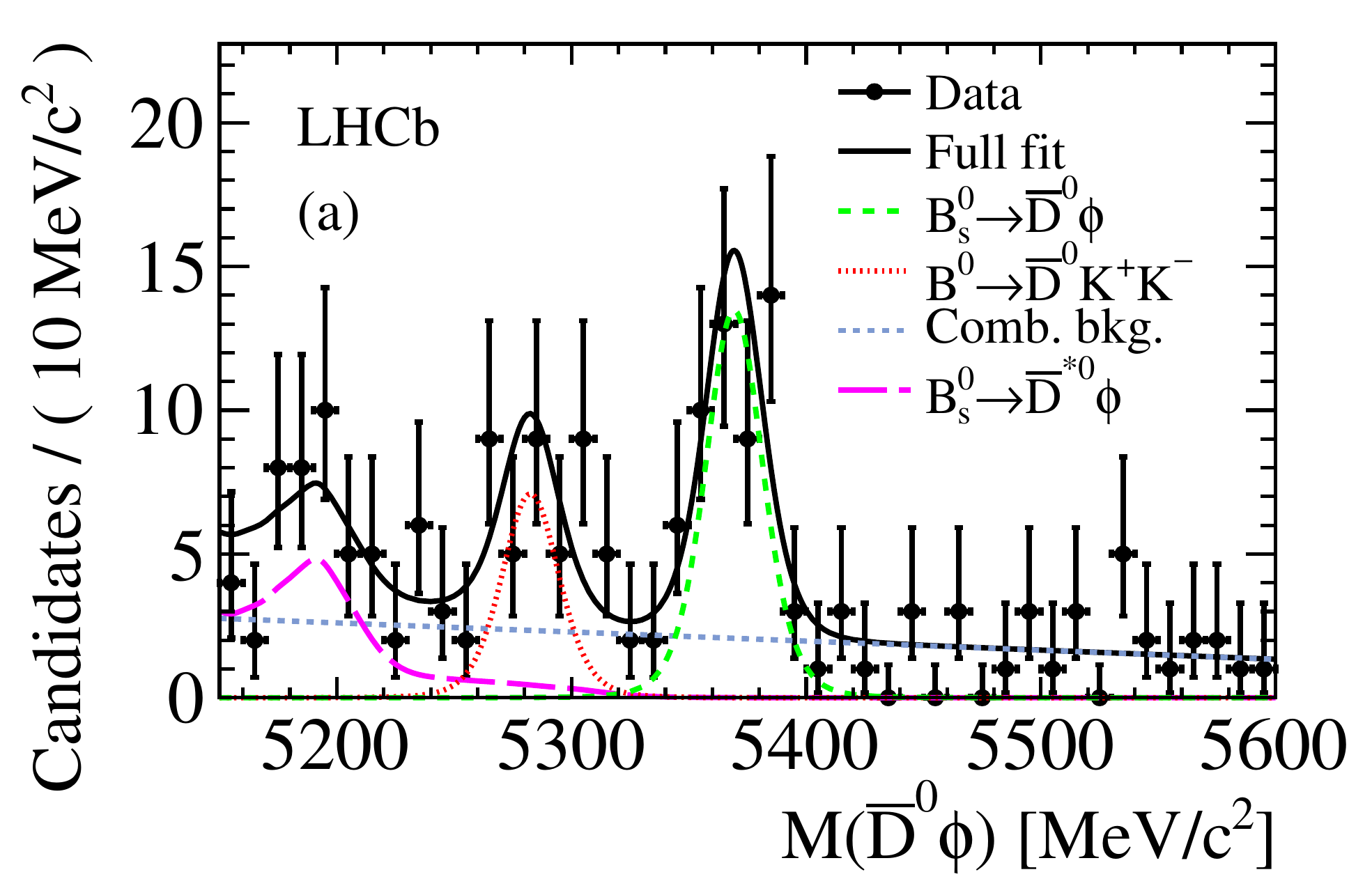}
    \includegraphics[width=0.49\linewidth]{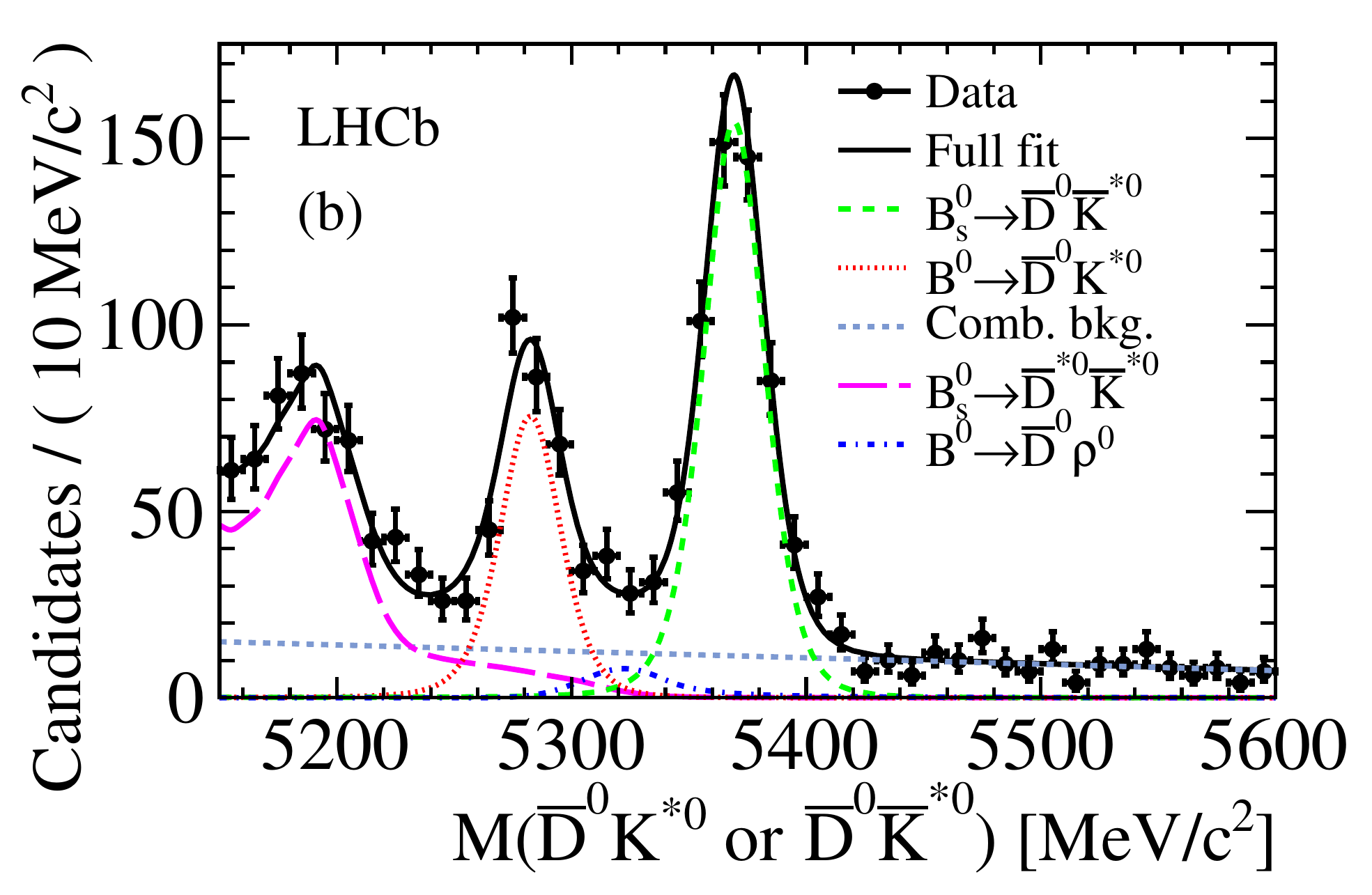}
  \end{center}
  \caption{
    \small 
Invariant mass distributions for (a) \BsToDphi, and (b) $\Bz\to\Dzb\Kstarz$ or $\Bs\to\Dzb\Kstarzb$ decays.
Data points are shown in black, the total fitted PDF as solid black line, 
and the components as detailed in the legends.
    }
  \label{fig:fit}
\end{figure}

Signal yield ratios are directly determined in the fit to take into account statistical correlations in 
the measurement of ratios of branching fractions. 
In total, there are 13 free parameters in the fit, including the background yields of the different components and the overall normalisation. The invariant mass distributions with the resulting fits are shown in Fig.~\ref{fig:fit}. 
 
\begin{figure}[tb!]
  \begin{center}
    \includegraphics[height=0.49\linewidth]{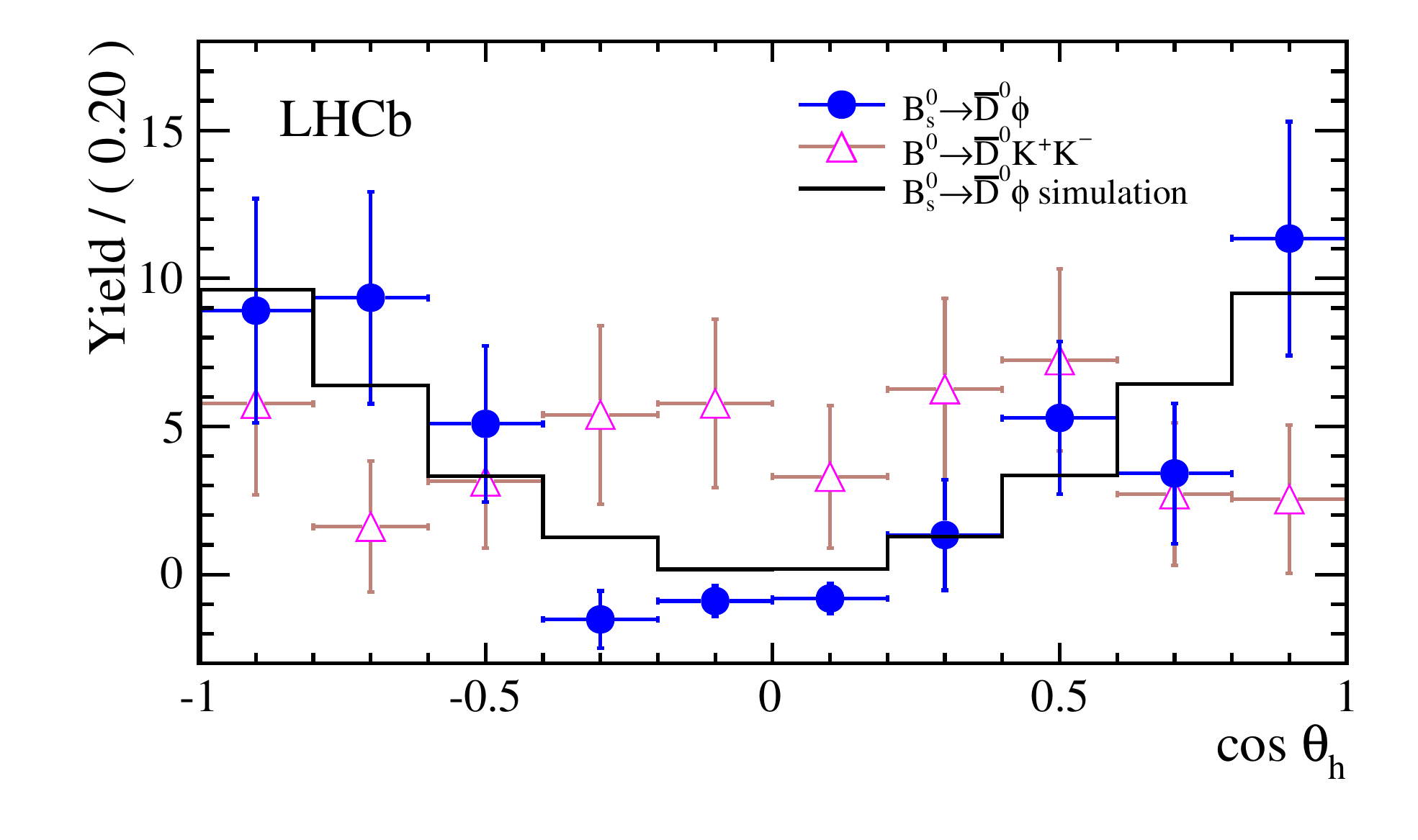}
    \vspace*{-1.0cm}
  \end{center}
  \caption{
    \small 
Distribution of the cosine of the helicity angle of the $\phi$ candidates. 
    }
  \label{fig:helicity}
\end{figure}

The helicity angle distribution of the $\phi$ candidates for the \Bs and \Bz signal is investigated.
The \sPlot~\cite{bib:sPlot} technique is adopted to assign a weight to the events and determine the signal components, using
the $\Dzb\phi$ invariant mass as the discriminating variable. 
For this purpose, the requirement on $\mathrm{cos}\,\theta_h >0.4$ has been lifted prior to the computation of the signal weights. 
The data distributions of $\mathrm{cos}\,\theta_h$, shown in Fig.~\ref{fig:helicity}, 
are compared to the expected distribution of \BsToDphi decays from simulation. 
The distribution observed for the $\Bd\to\Dzb\Kp\Km$ decay candidates is consistent 
with the expectation that this decay is not dominated by a pseudoscalar-vector quasi-two-body final state. 

\begin{figure}[tb!]
  \begin{center}
    \includegraphics[width=0.49\linewidth]{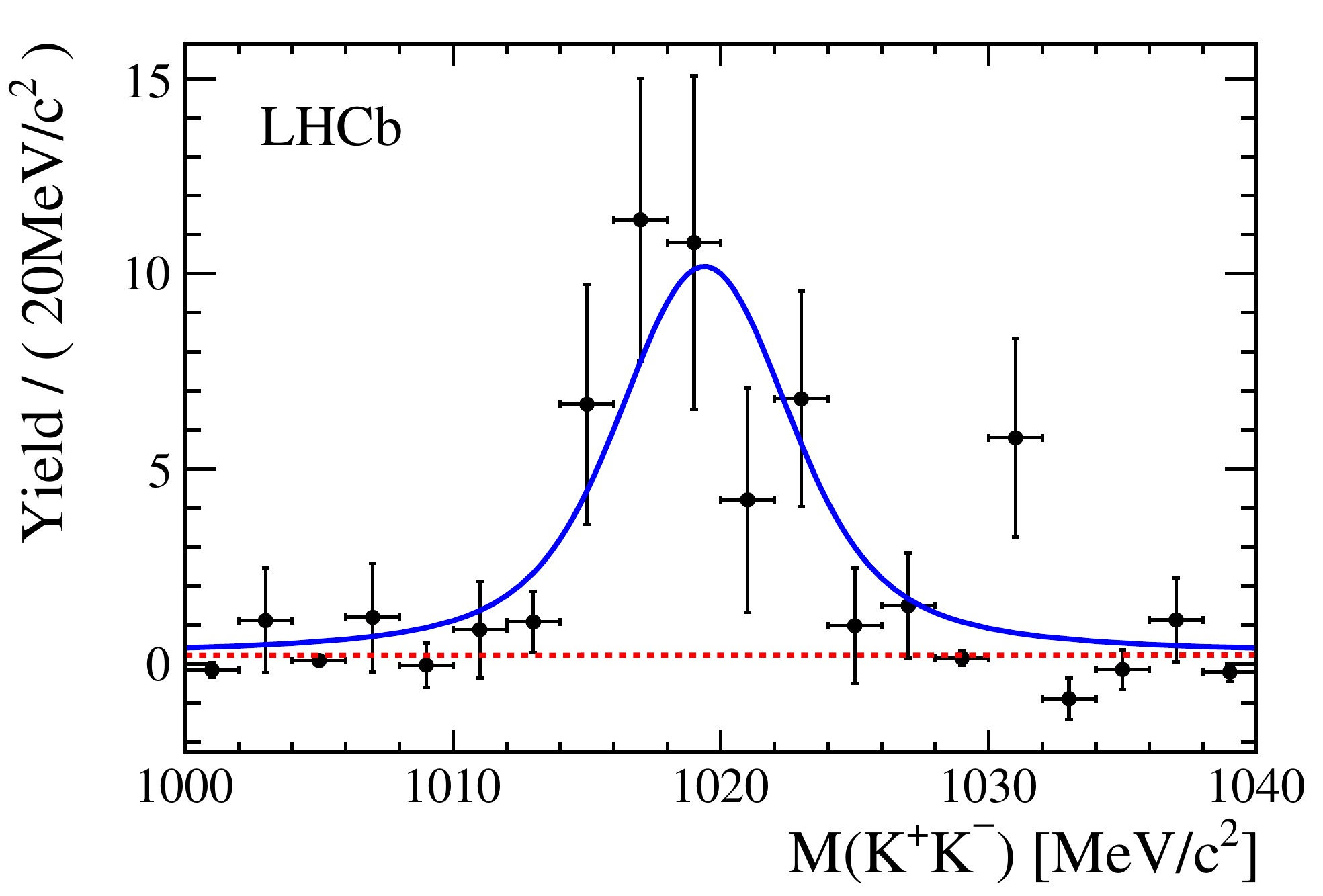}
    \includegraphics[width=0.49\linewidth]{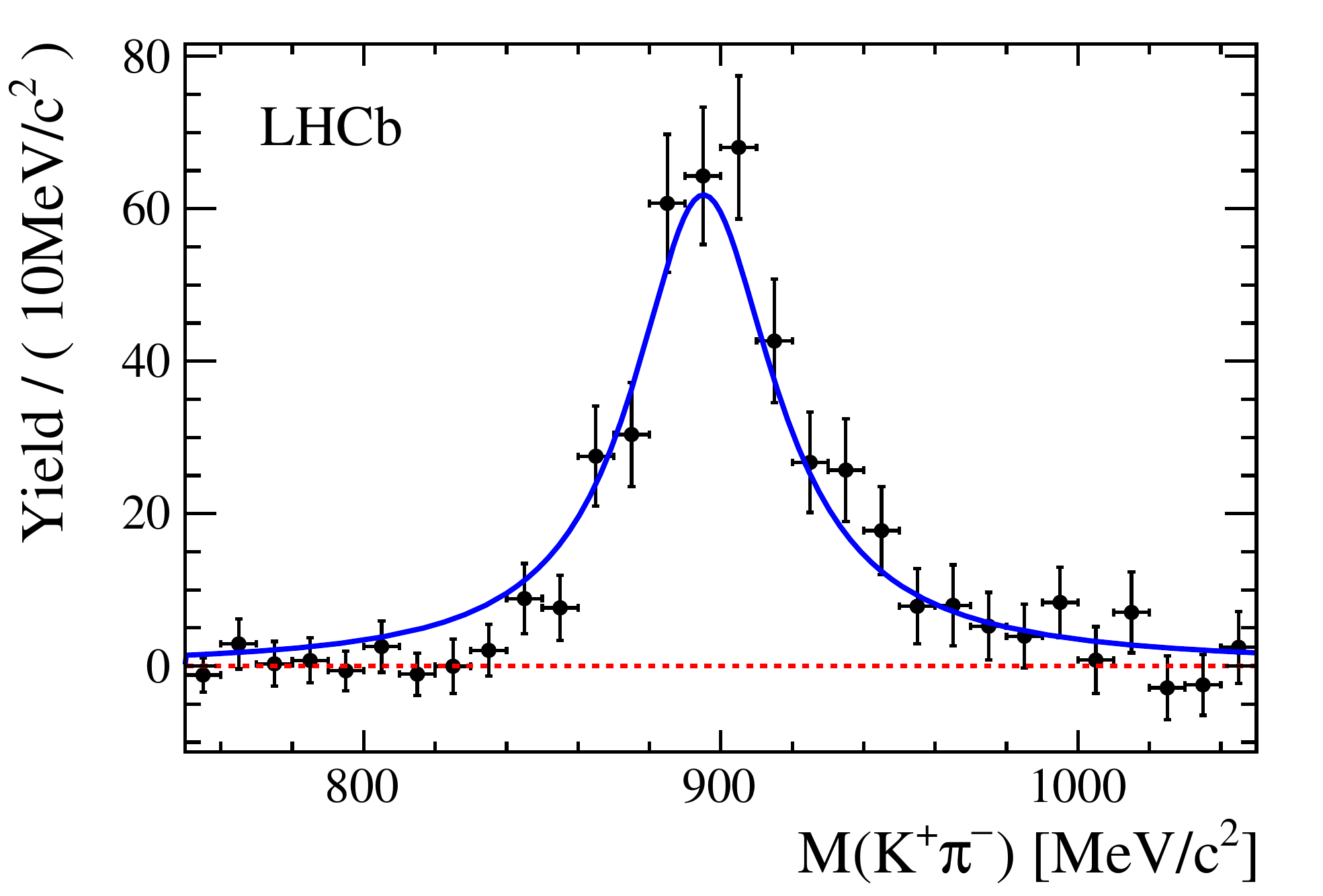}
  \end{center}
  \caption{
    \small
Background-subtracted distributions of the reconstructed (left) \Pphi mass from the \BsToDphi decay and
(right) $\Kstarzb$ mass from the \BsToDKstar decay. The dashed red line represents the S-wave component,
the solid blue line the total fit result.
    }
  \label{fig:kstar}
\end{figure}

The signal yield ratios are corrected for two residual backgrounds that peak at the mass of the \Bs or \Bz meson and are distributed as the signal. The first of the two backgrounds is the charmless background due to the decays $\Bs\to\Kp\pim\Kstarzb$ and $\Bz\to\Kp\pim\Kstarz$ proceeding without the presence of an intermediate \Dzb meson. There is no evidence of such background in the \BsToDphi channel.
 A large fraction of the charmless background in the $\Dzb\Kstarz$ final state is rejected with the requirement of a 
minimal \Dz flight distance introduced in Sec.~\ref{sec:Selection}.
 The remaining charmless background is evaluated using candidates from the \Dz sidebands. The $B$ yields in the \Dz sidebands above a linear background are extrapolated to the \Dz signal region and used to correct the signal. The uncorrected signal yields and the background contributions are given in Table~\ref{tab:yields}.
\begin{table}[tb]
  \caption{
    \small 
    Uncorrected signal yields and the peaking (charmless, S-wave) background yields.  
    }
\begin{center}\begin{tabular}{lccc}
    Channel                        & Signal & Charmless background & S-wave background\\\hline
\rule{0pt}{3ex}$\BsToDphi$                & 43$\,\pm\,$8      & $0\,\pm\,2$ & 2$\,\pm\,3$\\
    $\BsToDKstar$              & 535$\,\pm\,$30    & $4\,\pm\,3$ & 24$\,\pm\,$7   \\
    $\BdToDKstar$              & 260$\,\pm\,$24    & $4\,\pm\,3$ & 13$\,\pm\,$6   \\
  \end{tabular}\end{center}
\label{tab:yields}
\end{table}
The other source of peaking background is due to higher mass resonances and non-resonant  $\Bs\to\Dzb\Kp\Km$, $\Bs\to\Dzb\Km\pip$, and 
$\Bz\to\Dzb\Kp\pim$ decays that fall in the \BsToDphi, \BsToDKstar, and \BdToDKstar signal regions, 
respectively. This contribution is evaluated with fits to the \Pphi and \Kstarz background-subtracted  mass distributions in a wider range than the signal window. The background subtraction is performed using the \sPlot\ technique, 
with the $\Dzb\phi$ and $\Dzb\Kstarzb$ (or $\Dzb\Kstarz$) mass as discriminating variables. 
A linear PDF describes the S-wave background in the $\Dzb\Pphi$ final state. 
A spin-one Breit-Wigner distribution convolved with a Gaussian resolution function describes the signal, and an S-wave PDF the non-resonant background. 
The S-wave component in the \BdToDKstar and \BsToDKstar channels 
takes into account non-resonant and $K^{*0}(1430)$ resonance contributions and uses experimental input from the LASS experiment~\cite{bib:LASS,*bib:LASS2}. It is approximately linear in the region of interest, $\pm\,200\mevcc$ around the \Kstarz nominal mass. Potential interference effects between the S-wave and the P-wave components are covered by the assigned systematic uncertainty. 
The \Pphi and $\Kstarzb$ mass distributions are shown in Fig.~\ref{fig:kstar}.
The background yields, after extrapolation to the  \Kstarz and \Pphi signal mass windows, are listed in Table~\ref{tab:yields}.

A likelihood ratio test is employed to assess the statistical significance of the 
\BsToDphi signal, which is given by $\sqrt{2{\rm ln}(\mathcal{L}_{\rm s+b}/\mathcal{L}_{\rm b})}$ and found to be 7.1 standard deviations. Here $\mathcal{L}_{\rm s+b}$ and $\mathcal{L}_{\rm b}$
are the maximum values of the likelihoods for the signal-plus-background and background-only hypotheses, respectively. 

The ratios of branching fractions 
are evaluated from the uncorrected signal yields, $N$, and the sum of the charmless and non-resonant background yields,
$N^{\rm bkg}$, as
\begin{equation}
\label{eq:BRDphi}
{\cal{R}}_{\phi} \equiv \frac{{\cal B}(\BsToDphi)}{{\cal{B}}(\BsToDKstar)} = 
\frac{N_{\BsToDphi}}{N_{\BsToDKstar}} \cdot
\frac{\left(1-\frac{N^{\rm bkg}_{\BsToDphi}}{N_{\BsToDphi}}\right)}
{\left(1-\frac{N^{\rm bkg}_{\BsToDKstar}}{N_{\BsToDKstar}}\right)}\cdot
\frac{\epsilon_{\BsToDKstar}}{\epsilon_{\BsToDphi}}\cdot
\frac{{\cal{B}}(\Kstarz\to\Kp\pim)}{{\cal{B}}(\Pphi\to\Kp\Km)},
\end{equation}
and
\begin{equation}
\label{eq:BRDKstar}
{\cal{R}}_{\Kstarz} \equiv \frac{{\cal B}(\BsToDKstar)}{{\cal{B}}(\BdToDKstar)} = 
\frac{N_{\BsToDKstar}}{N_{\BdToDKstar}} \cdot
\frac{\left(1-\frac{N^{\rm bkg}_{\BsToDKstar}}{N_{\BsToDKstar}}\right)}
{\left(1-\frac{N^{\rm bkg}_{\BdToDKstar}}{N_{\BdToDKstar}}\right)}\cdot
\frac{\epsilon_{\BdToDKstar}}{\epsilon_{\BsToDKstar}}\cdot
\left(\frac{f_s}{f_d}\right)^{-1},
\end{equation}
where the ratio of the \Bs and \Bz fragmentation fractions is  $f_s/f_d = 0.256\,\pm\,0.020$~\cite{bib:fsfd}, the value of the $\phi\to \Kp\Km$ branching fraction is 0.489$\,\pm\,$0.005~\cite{PDG2012}, and ${\cal{B}}(\Kstarz\to\Kp\pim) = 2/3$.
The total efficiencies, $\epsilon$,  account for the geometrical acceptance of the detector, the reconstruction, the event selection, the PID, and the trigger efficiencies. All efficiencies are computed from simulated events, except for the PID and hardware trigger efficiencies, which are obtained from data, using a high-purity calibration sample of $\Dstarp\to \Dz(\to\Km\pip)\pip$ decays.
The resulting ratios of branching fractions are 
${\cal{R}}_{\phi} =  0.069 \pm 0.013$ and ${\cal{R}}_{\Kstarz} = 7.8 \pm 0.7$, where the uncertainties are statistical only.

\section{Systematic uncertainties}
\label{sec:Systematics}

Several sources of systematic uncertainties are considered.
Those associated to the trigger and PID selection affect only ${\cal{R}}_\phi$ and are mainly due to systematic uncertainties in the calibration procedure. The ratios of the efficiencies of the decays $\BsToDphi$ and $\BsToDKstar$ for the trigger and PID are found to be 0.97$\,\pm\,$0.05 and 1.08$\,\pm\,$0.03, respectively, where the errors are propagated as systematic uncertainties to  ${\cal{R}}_\phi$.

Similarly, the uncertainty on the efficiencies of the charm meson flight distance selection affects only ${\cal{R}}_\phi$, where different criteria are chosen for the $\BsToDphi$ and $\BsToDKstar$ modes. The ratio of the corresponding efficiencies is found to be 1.27$\,\pm\,$0.03, where the uncertainty includes a contribution from the difference between data and simulation. In order to estimate the efficiency 
in data, the fit to the invariant mass of the $B$ candidates is performed to data samples selected with all criteria except that on the flight distance. 
For this sample, the charmless background contribution is estimated 
using events in the upper $D$ mass sideband and subtracted from the signal yields. 

The ratio of the efficiencies for the decays $\BsToDphi$ and $\BsToDKstar$ of the remaining selection criteria is found to be $1.21\,\pm\,0.03$, where the deviation from unity is mainly due to the different widths and mass windows for the $\phi$ and $\Kstarz$ resonances. The ratio of the efficiencies for the decays $\BsToDKstar$ and $\BdToDKstar$ is found from simulation to be $1.04\,\pm\,0.01$. The uncertainties on these efficiencies are propagated as systematic uncertainties due to the selection.

The fit procedure is validated using simulated pseudo-experiments. 
The fit bias, relative to the fitted ratio, is evaluated to be 1.4\% for ${\cal{R}}_\phi$ and 0.2\% for ${\cal{R}}_{\Kstarz}$ and is assigned as systematic uncertainty. 
The signal model uncertainty is evaluated by varying the fixed signal parameters by 10\%, 
which is about three times the difference between data and simulation, as determined by a fit where those parameters are free to vary.
The background shape uncertainty is determined from the bias in the results obtained by fitting samples generated with an alternative (exponential) combinatorial background model.

The uncertainties on the charmless background yields given in Table~\ref{tab:yields} are assumed to be uncorrelated and are propagated to assign the associated systematic uncertainty. Similarly, the statistical uncertainties on the S-wave background yields are propagated to ${\cal{R}}_\phi$ and  ${\cal{R}}_{\Kstarz}$ to assign respective systematic uncertainties due to the non-resonant correction.

A summary of the systematic uncertainties is given in Table~\ref{tab:systematics}. The uncertainty on the fragmentation fraction
$f_s/f_d$, which is the dominant systematic uncertainty for ${\cal{R}}_{\Kstarz}$, is not included, and
is listed separately.
\begin{table}[t]
  \caption{
    \small 
    Absolute systematic uncertainties of the measured ratio of branching fractions. The total is obtained as sum in quadrature of the different contributions.
    }
\begin{center}\begin{tabular}{lcc}
    Source                    &  ${\cal{R}}_\phi$    & ${\cal{R}}_{\Kstarz}$ \\\hline
\rule{0pt}{3ex}Trigger                 &0.003 & -    \\  
    PID                     &0.002 & -     \\ 
    Flight distance         &0.002 & -     \\ 
    Selection                &   0.002& - \\ 
    Simulation statistics     &  0.001& 0.10 \\
    Fit bias                   & 0.001& 0.03  \\
    Signal model                &0.001& 0.04  \\
    Background model            &0.001& 0.01  \\
    Charmless correction         &   0.003& 0.10  \\
    Non-resonant correction      &   0.004& 0.22  \\
    \Pphi branching fraction     &   0.001& - \\ \hline
\rule{0pt}{1ex}    Total                        &   0.007& 0.26  \\
  \end{tabular}\end{center}
\label{tab:systematics}
\end{table}

\section{Results and conclusions}
\label{sec:Results}

The significance of the \BsToDphi signal, including systematic uncertainties, is obtained by scaling the statistical significance with the ratio of the statistical to the total (statistical and systematic) uncertainty on the signal yield. It is found to be
6.5 standard deviations. This decay is therefore observed for the first time.

The ratios of branching fractions are found to be
\[{\cal{R}}_{\phi} = 0.069 \pm 0.013 ~(\rm stat) \pm 0.007 ~(\rm syst),\]
\[{\cal{R}}_{\Kstarz} = 7.8 \pm 0.7 ~({\rm stat}) \pm 0.3 ~({\rm syst}) \pm 0.6 ~(f_s/f_d).\]
From ${\cal{R}}_{\Kstarz}$ and 
the value of the \BdToDKstar branching fraction from Ref.~\cite{PDG2012},
the \BsToDKstar branching fraction is calculated to be
\[{\cal{B}}(\BsToDKstar) = [3.3 \pm 0.3 ~({\rm stat}) \pm 0.1 ~({\rm syst}) \pm 0.3 ~(f_s/f_d) \pm 0.5 ~({\cal{B}}(\BdToDKstar))] \times 10^{-4} .\]
This result is consistent with and improves on the previous determination by \lhcb ~\cite{bib:LHCb-PAPER-2011-008}, which is based on an independent data sample.
Using the above results for ${\cal{R}}_{\phi}$, ${\cal{R}}_{\Kstarz}$ and the \BdToDKstar branching fraction, the branching fraction for
\BsToDphi is calculated to be
\[{\cal{B}}(\BsToDphi) =  [2.3 \pm 0.4 ~({\rm stat}) \pm 0.2 ~({\rm syst})  \pm 0.2 ~(f_s/f_d) \pm 0.3 ~({\cal{B}}(\BdToDKstar))] \times 10^{-5}, \]
which takes into account the correlation in the statistical uncertainties between ${\cal{R}}_{\phi}$ and ${\cal{R}}_{\Kstarz}$ of $-13.6$\%.
The correlation between the corresponding systematic uncertainties is negligible.
The central value is about a factor two smaller than the branching fraction for the \BdToDKstar decay
and supports the observation of SU(3) breaking effects in other colour suppressed 
$B^0_{(s)}\to \Dzb V$ decays~\cite{bib:LHCb-PAPER-2011-008}, where $V$ is a vector meson. 
With larger data samples, the \BsToDphi decay will contribute to the measurements of the \CP violating phases $\gamma$ and $\beta_s$.

\section*{Acknowledgements}

\noindent We express our gratitude to our colleagues in the CERN
accelerator departments for the excellent performance of the LHC. We
thank the technical and administrative staff at the LHCb
institutes. We acknowledge support from CERN and from the national
agencies: CAPES, CNPq, FAPERJ and FINEP (Brazil); NSFC (China);
CNRS/IN2P3 and Region Auvergne (France); BMBF, DFG, HGF and MPG
(Germany); SFI (Ireland); INFN (Italy); FOM and NWO (The Netherlands);
SCSR (Poland); MEN/IFA (Romania); MinES, Rosatom, RFBR and NRC
``Kurchatov Institute'' (Russia); MinECo, XuntaGal and GENCAT (Spain);
SNSF and SER (Switzerland); NAS Ukraine (Ukraine); STFC (United
Kingdom); NSF (USA). We also acknowledge the support received from the
ERC under FP7. The Tier1 computing centres are supported by IN2P3
(France), KIT and BMBF (Germany), INFN (Italy), NWO and SURF (The
Netherlands), PIC (Spain), GridPP (United Kingdom). We are thankful
for the computing resources put at our disposal by Yandex LLC
(Russia), as well as to the communities behind the multiple open
source software packages that we depend on.

\addcontentsline{toc}{section}{References}
\setboolean{inbibliography}{true}
\bibliographystyle{LHCb}
\bibliography{main,LHCb-PAPER,LHCb-CONF,LHCb-DP}

\end{document}